\documentclass[runningheads]{llncs}
\usepackage[T1]{fontenc}
\usepackage{graphicx}

\usepackage{amsmath} 
\usepackage[htt]{hyphenat}
\usepackage{multirow} 
\usepackage{multicol}

\usepackage{lipsum}

\newcommand\blfootnote[1]{%
  \begingroup
  \renewcommand\thefootnote{}\footnote{#1}%
  \addtocounter{footnote}{-1}%
  \endgroup
}
\begin{document}
\title{Exploiting temporal parallelism for LSTM Autoencoder acceleration on FPGA}
%
%

\author{Aimilios Leftheriotis\inst{1} \and
Dimosthenis Masouros\inst{2}\and
Dimitrios Soudris\inst{2}\and
George Theodoridis\inst{1}}
\authorrunning{A. Leftheriotis et al.}
\institute{University of Patras, Patras, Greece\\
\email{aleftheriotis@ac.upatras.gr, theodor@upatras.gr} \and
National Technical University of Athens, Athens, Greece\\
\email{dmasouros@microlab.ntua.gr, dsoudris@microlab.ntua.gr}}
\maketitle              
\begin{abstract}
Recurrent Neural Networks (RNNs) are vital for sequential data processing. 
Long Short-Term Memory Autoencoders (LSTM-AEs) are particularly effective for unsupervised anomaly detection in time-series data. 
However, inherent sequential dependencies limit parallel computation.
While previous work has explored FPGA-based acceleration for LSTM networks, efforts have typically focused on optimizing a single LSTM layer at a time. 
We introduce a novel FPGA-based accelerator using a dataflow architecture that exploits \textit{temporal parallelism} for concurrent multi-layer processing of different timesteps within sequences. 
Experimental evaluations on four representative LSTM-AE models with varying widths and depths, implemented on a Zynq UltraScale+ MPSoC FPGA, demonstrate significant advantages over CPU (Intel Xeon Gold 5218R) and GPU (NVIDIA V100) implementations.
Our accelerator achieves latency speedups up to 79.6x vs. CPU and 18.2x vs. GPU, alongside energy-per-timestep reductions of up to 1722x vs. CPU and 59.3x vs. GPU.
These results, including superior network depth scalability, highlight our approach's potential for high-performance, real-time, power-efficient LSTM-AE-based anomaly detection on FPGAs.

\end{abstract}
\section{Introduction}
\label{sec:Introduction}

Recurrent Neural Networks (RNNs) and especially Long Short-Term Memory (LSTM) networks~\cite{hochreiter1997long}, are crucial for sequential data processing due to their proficiency in managing long-term dependencies. 
LSTM-Autoencoders (LSTM-AEs), excel at unsupervised anomaly detection in time-series data.
By training on benign data, LSTM-AEs learn an efficient encoding of typical input distributions and are widely applied in domains like network traffic monitoring, arrhythmia detection, and abnormal gait recognition~\cite{kieu2018outlier,liu2022arrhythmia,jun2020feature}.

Real-time inference for RNNs on large datasets demands high-performance, energy-efficient accelerators.
However, recurrent data dependencies (intra- and inter-sequence) limit parallelism on CPUs and GPUs. 
FPGAs offer reconfigurability, energy efficiency, and better dependency management. 
Prior FPGA LSTM accelerators~\cite{ferreira2016fpga,liang2023f,guan2017fpga} often optimize a single timestep/layer. 
SHARP~\cite{aminabadi2023sharp} pipelines multiple timesteps but still processes one layer at a time.

LSTM-AEs feature encoders that compress input features and decoders that reconstruct them, using multiple LSTM layers of varying, often small, sizes. 
Consequently, traditional layer-by-layer acceleration underutilizes hardware for smaller layers. 
An accelerator exploiting \textit{temporal parallelism} by distributing computation across multiple LSTM layers concurrently would improve efficiency.

In this paper (i) we introduce an open-source\footnote{\url{https://github.com/aimilefth/samos25\_lstm\_ae\_fpga}}\blfootnote{This work was partially funded by the PRIVATEER project, supported by the Smart Networks and Services Joint Undertaking (SNS JU) under the European Union’s Horizon Europe programme (Grant Agreement No. 101096110)} FPGA-based LSTM-AE accelerator enabling concurrent processing of multiple LSTM layers across different timesteps, exploiting temporal parallelism, a novel approach in this domain.
(ii) We also present a methodology for dataflow balancing using reuse factor configurations to equalize per-timestep latency across LSTM modules, thereby minimizing resource underutilization and enhancing throughput. 
(iii) Furthermore, we implement and evaluate our accelerator on four representative LSTM-AE models on a ZCU104 MPSoC FPGA, achieving significant performance and energy efficiency improvements over CPU and GPU platforms, achieving latency speedups up to 79.6x (vs. CPU) and 18.2x (vs. GPU), alongside energy-per-timestep reductions up to 1722x (vs. CPU) and 59.3x (vs. GPU).


The remainder of this paper is organized as follows.
Section \ref{sec:Background} provides background on LSTMs and LSTM-AEs.
Section \ref{sec:Proposed Accelerator} details the proposed accelerator.
Section \ref{sec:Evaluation} presents the evaluation.
Finally, Section \ref{sec:Conclusion} concludes the paper.
\section{Background}
\label{sec:Background}

\begin{figure}[htpb]
    \centering
    \includegraphics[width=0.60\columnwidth]{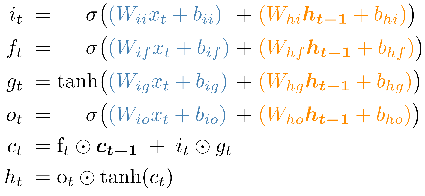}
    \caption{LSTM equations. The two MVMs are colored, the inter-sequence dependencies are in bold.}
    \label{fig:LSTM_equation}
    \vspace{-10pt}
\end{figure}

LSTM networks, a class of RNNs, excel at sequence modeling by capturing long-term dependencies.
Figure \ref{fig:LSTM_equation} illustrates the computations an LSTM layer performs at each timestep $t$. 
The LSTM processes the current input $\mathbf{x}_t$ and the previous hidden state $\mathbf{h}_{t-1}$, computing four internal gate vectors ($\mathbf{i}_t, \mathbf{f}_t, \mathbf{g}_t, \mathbf{o}_t$) that regulate the memory cell state $\mathbf{c}_t$. 
The output is the updated hidden state $\mathbf{h}_t$. 
Central to these are two Matrix-Vector Multiplications (MVMs), one on $\mathbf{x}_t$ and one on $\mathbf{h}_{t-1}$ (highlighted in Figure \ref{fig:LSTM_equation}), forming most of the computational load and being highly parallelizable. 
However, intra-sequence dependencies (between $\mathbf{c}_t, \mathbf{h}_t$) and inter-sequence dependencies (on $\mathbf{c}_{t-1}, \mathbf{h}_{t-1}$) constrain parallelization.

LSTM-Autoencoders (LSTM-AEs) use LSTMs in both the encoder and the decoder parts. 
The encoder compresses input data into a latent representation, and the decoder reconstructs the original input. 
This is effective for sequential data, allowing LSTM-AEs to learn patterns in multivariate time-series. 
Trained on benign data, they learn and overfit on normal behavior.
On the contrary, anomalous sequences lead to higher reconstruction error, indicating anomalies.
\section{Proposed Accelerator}
\label{sec:Proposed Accelerator}

\subsection{Architecture Overview}
\begin{figure}[htbp]
    \centering
    \includegraphics[width=0.99\columnwidth]{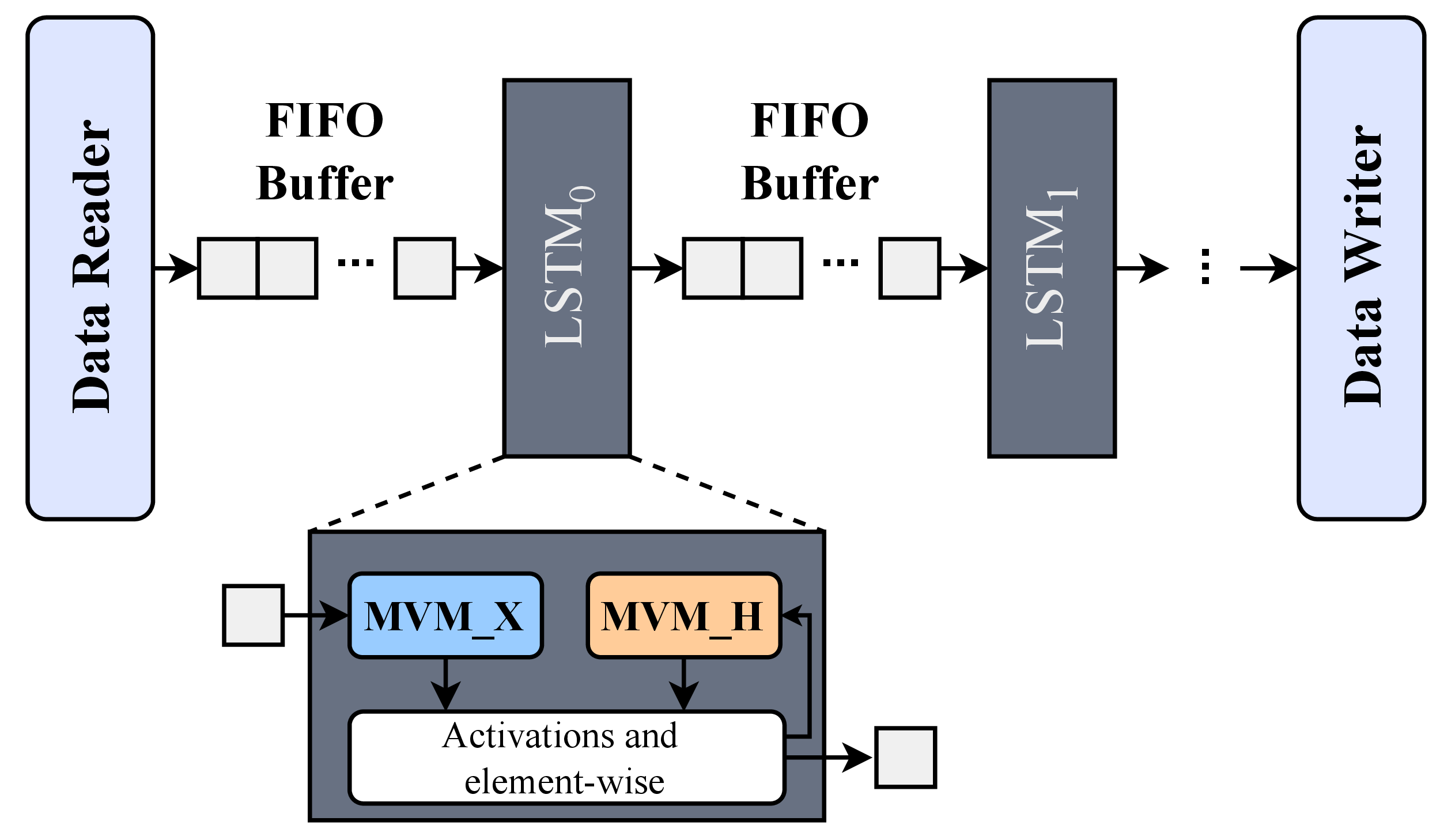}
    \caption{High-level overview of our dataflow architecture. MVM\_X and MVM\_H execute the blue and orange computations of Figure \ref{fig:LSTM_equation}, respectively}
    \label{fig:Architecture}
    \vspace{-15pt}
\end{figure}


The proposed LSTM-AE FPGA-based accelerator is depicted in Figure \ref{fig:Architecture}. 
The architecture leverages a dataflow paradigm, enabling concurrent execution of all processing modules and inter-module communication exclusively through FIFO queues. 
Each LSTM layer is mapped to a corresponding LSTM$_i$ module. 
The accelerator interfaces with external DRAM through two dedicated modules: the \textit{Data Reader}, which streams input sequences from memory, and the \textit{Data Writer}, which stores the processed outputs back into DRAM.

Each LSTM$_i$ module has a dataflow microarchitecture with \texttt{MVM\_X}, \texttt{MVM\_H} units for MVMs, and an \textit{Activations and Element-Wise unit}. These also use FIFOs. MVM units process input elements sequentially, parallelizing multiply-accumulates. Processed timesteps are forwarded for pipelined activation and element-wise computations.

\subsection{Execution Scheduling and Latency Estimation}

The dataflow architecture enables temporal parallelism.
Once the initial pipeline fill latency has elapsed, each \texttt{LSTM$_i$} module operates concurrently, processing a different timestep of the input sequence ($t_j$, $t_{j+1}$, $t_{j+2}$, etc.) relative to its adjacent modules. 
This staggered, pipelined execution, where computations for different timesteps overlap across layers, creates a continuous processing flow, enhancing throughput compared to layer-by-layer execution.

The total latency of the accelerator, $Acc\_Lat$, in clock cycles, to process a sequence of length $T$ with $N$ LSTM layers is:

\begin{equation}
Acc\_Lat = T\cdot Lat\_t_m + \sum_{i=0}^{m-1}(Lat\_t_i) + \sum_{i=m+1}^{N-1}(Lat\_t_i)
\label{eq:Acc_Lat}
\end{equation}
where $Lat\_t_i$ represents the per-timestep latency of \texttt{LSTM$_i$} and  \linebreak$Lat\_t_m = \max_{i}(Lat\_t_i)$ is the maximum per-timestep latency across all $N$ layers, representing the pipeline's bottleneck stage. 
$Lat\_t_i$ is derived by:
\begin{equation}
    Lat\_t_i = \max(X\_t_i,H\_t_i)   
\label{eq:Lat_t_i}
\end{equation}

where $X\_t_i$ and $H\_t_i$ are latencies of \texttt{MVM\_X$_i$} and \texttt{MVM\_H$_i$}.
Assuming an \texttt{LSTM$_i$} module has an input feature dimension $LX_i$ and a hidden state dimension $LH_i$, these MVM latencies depend on the dimensions and the configured hardware reuse factors, $RX_i$ and $RH_i$:

\vspace{-1ex}
\noindent\begin{multicols}{2}
 \noindent\begin{equation}
      X\_t_i = LX_i\cdot RX_i + LH_i
      \label{eq:X_t_i}
  \end{equation}
  \begin{equation}
      H\_t_i = LH_i\cdot RH_i + LH_i
      \label{eq:H_t_i}
  \end{equation}
\end{multicols}



\vspace{-1ex}
Hardware reuse factors $RX_i, RH_i$ are cycles per input/hidden state element, inversely proportional to parallel multipliers $MX_i, MH_i$:
\vspace{-1ex}
\noindent\begin{multicols}{2}
 \noindent\begin{equation}
    RX_i = \frac{4 \cdot LH_i}{MX_i} 
\label{eq:RX_i}
  \end{equation}
  \begin{equation}
    RH_i = \frac{4 \cdot LH_i}{MH_i} 
\label{eq:RH_i}
  \end{equation}
\end{multicols}


The term $4 \cdot LH_i$ represents the total number of multiply-accumulate operations required to process one element of the input vector ($\mathbf{x}_t$) or hidden state vector ($\mathbf{h}_{t-1}$) for the corresponding MVM across the four LSTM gates.


\subsection{Dataflow Balancing}
\label{subsec:Dataflow Balancing}

Achieving a balanced dataflow, where \texttt{LSTM$_i$} modules have similar per-timestep latencies ($Lat\_t_i$), is critical for maximizing efficiency. 
An imbalanced pipeline, where one module is significantly slower, creates a bottleneck, causing faster upstream modules to stall, underutilizing resources and limiting throughput.
To mitigate imbalances, the reuse factors are configured. 
Within each \texttt{LSTM$_i$}, MVM units should have equal latency ($X\_t_i = H\_t_i$) to avoid idling, leading to:

\begin{equation}
    RX_i = \frac{LH_i}{LX_i}\cdot RH_i
\label{eq:RX_i,RH_i}
\end{equation}

Secondly, to equalize $Lat\_t_i$ across all modules to match the bottleneck latency $Lat\_t_m$ (dominated by $H\_t_m$ if internally balanced), $RH_i$ is derived relative to $RH_m$:


\begin{equation}
    RH_i = \frac{LH_m - LH_i}{LH_i} + \frac{LH_m}{LH_i} RH_m
\label{eq:RH_i,RH_m}
\end{equation}

Utilizing Equations \eqref{eq:RX_i,RH_i} and \eqref{eq:RH_i,RH_m} promotes a balanced dataflow. Determining the optimal $RH_m$ for a given model and platform is future work.


\subsection{Benefits of Dataflow and Temporal Parallelism}

Temporal parallelism addresses RNN challenges like inter/intra-sequence dependencies limiting conventional hardware.
Our design diverges from traditional layer-by-layer execution, which underutilizes hardware, especially for smaller LSTM layers common in LSTM-AEs.

The dataflow architecture offers critical advantages: concurrent module operation minimizes idle times, and localized communication via FIFOs reduces global memory access, mitigating bandwidth bottlenecks and timing issues.
The modular dataflow approach facilitates scalability, since adding or modifying LSTM layers minimally impacts the core architecture.
It also supports fine-grained resource optimization for high utilization across network sizes.

\section{Evaluation}
\label{sec:Evaluation}

\subsection{Experimental Setup}
\label{subsec:ExperimentalSetup}


To evaluate our accelerator, we implemented four LSTM-AE models: \texttt{LSTM-AE-F32-D2}, \texttt{LSTM-AE-F32-D6}, \texttt{LSTM-AE-F64-D2}, and \texttt{LSTM-AE-F64-D6}. 
The naming \texttt{LSTM-AE-F\{$X$\}-D\{$Y$\}} indicates an input feature size of $X$ and $Y$ total LSTM layers (half encoder, half decoder, with feature sizes halving/doubling symmetrically from input/bottleneck). 
For example, \texttt{LSTM-AE-F32-D2} has 2 layers (32$\rightarrow$16$\rightarrow$32 features), and \texttt{LSTM-AE-F32-D6} has 6 layers (32$\rightarrow$16$\rightarrow$8$\rightarrow$4$\rightarrow$8$\rightarrow$16$\rightarrow$32 features). 
This selection facilitates a comprehensive performance assessment across a range of model complexities, from shallower networks with narrower inputs (\texttt{LSTM-AE-F32-D2}) to deeper networks processing wider inputs (\texttt{LSTM-AE-F64-D6}), representing typical configurations found in anomaly detection use-cases.

Designs were implemented on AMD Zynq\texttrademark\ UltraScale+\texttrademark\ MPSoC ZCU104 using Vitis 2021.1 (HLS for kernel, Vitis for integration), targeting 300 MHz. 
We used 32-bit fixed-point with 24 fractional bits (Q8.24)  and Piecewise Linear Approximations for sigmoid and tanh functions.

\begin{table}[htbp]
\centering
\caption{FPGA Resource Utilization (\%) and Hardware Reuse Factor $RH_m$} 
\begin{tabular}{|c|c|c|c|c|c|} 
\hline
\textbf{Name} & \textbf{$RH_m$} & \textbf{LUT\%} & \textbf{FF\%} & \textbf{BRAM\%} & \textbf{DSP\%} \\ \hline
\texttt{LSTM-AE-F32-D2} & 1 & 26.11 & 12.87 & 39.74 & 34.72 \\ \hline
\texttt{LSTM-AE-F64-D2} & 4 & 43.04 & 18.52 & 77.08 & 18.06 \\ \hline
\texttt{LSTM-AE-F32-D6} & 1 & 42.47 & 16.89 & 69.39 & 48.15 \\ \hline
\texttt{LSTM-AE-F64-D6} & 8 & 69.27 & 24.19 & 59.94 & 16.67 \\ \hline
\end{tabular}
\vspace{-15pt}
\label{tab:ResourceUtilization}
\end{table}


The primary hardware reuse factor $RH_m$ was configured for each model as specified in Table \ref{tab:ResourceUtilization}.
These values were determined based on the resource constraints of the target XCZU7EV FPGA, ensuring synthesizability while attempting to maximize exploited parallelism where possible.
Subsequently, the reuse factors for all other layers were derived using the dataflow balancing methodology detailed in Section \ref{subsec:Dataflow Balancing}.
Narrower input models (\texttt{F32}) allowed $RH_m=1$, achieving minimal reuse for bottleneck MVM. 
Wider models (\texttt{F64}) required $RH_m=4$ or $8$ due to FPGA BRAM and LUT limits. 
Minimizing $RH_m$ (maximizing MVM parallelism) needs concurrent BRAM access for many weight elements, increasing BRAM port and LUT logic demands, which become bottlenecks.

Resource utilization increases with wider features and greater depth. 
However, LSTM-AE depth often introduces smaller intermediate layers. 
Our dataflow balancing assigns these layers higher reuse factors ($RX_i, RH_i$), demanding proportionally fewer resources per layer than widening initial layers. 
Thus, adding depth has a less pronounced resource impact than increasing input feature dimensions. 
This $RH_m$-based configurability and balancing approach shows potential for various FPGAs, including resource-constrained embedded devices.

\subsection{Performance Comparisons}

We compared the latency and energy-efficiency of our FPGA accelerator against CPU (Intel\textregistered\ Xeon\textregistered\ Gold 5218R CPU @2.10GHz) and GPU (NVIDIA\textregistered\ V100) PyTorch just-in-time compiled baselines. 
We evaluated the four LSTM-AE models across sequence lengths of 1 to 64 timesteps, covering real-time and throughput scenarios.

\begin{table}[htbp]
\centering
\caption{Inference Latency (ms)}
\label{tab:LatencyComparison}
\begin{tabular}{|c|c|c@{}c|c@{}c||c|c@{}c|c@{}c|}
\hline
\multicolumn{11}{|c|}{\textbf{Models D2}} \\ \hline
\multirow{2}{*}{\textbf{\begin{tabular}[c]{@{}c@{}}Time-\\Steps\end{tabular}}} & \multicolumn{5}{c||}{\textbf{\texttt{LSTM-AE-F32-D2}}} & \multicolumn{5}{c|}{\textbf{\texttt{LSTM-AE-F64-D2}}} \\
\cline{2-11} 
& \textbf{FPGA} & \multicolumn{2}{c|}{\textbf{CPU}} & \multicolumn{2}{c|}{\textbf{GPU}} & \textbf{FPGA} & \multicolumn{2}{c|}{\textbf{CPU}} & \multicolumn{2}{c|}{\textbf{GPU}} \\
\hline
 1  & \textbf{0.033} & 0.420 & (x12.7) & 0.275 & (x8.3)  & \textbf{0.038} & 0.414 & (x10.9) & 0.272 & (x7.2) \\ \hline
 2  & \textbf{0.036} & 0.479 & (x13.3) & 0.273 & (x7.6)  & \textbf{0.050} & 0.542 & (x10.8) & 0.273 & (x5.5) \\ \hline
 4  & \textbf{0.037} & 0.550 & (x14.9) & 0.269 & (x7.3)  & \textbf{0.059} & 0.613 & (x10.4) & 0.279 & (x4.7) \\ \hline
 6  & \textbf{0.038} & 0.591 & (x15.6) & 0.274 & (x7.2)  & \textbf{0.069} & 0.596 & (x8.6)  & 0.279 & (x4.0) \\ \hline
16  & \textbf{0.048} & 0.887 & (x18.5) & 0.288 & (x6.0)  & \textbf{0.118} & 0.923 & (x7.8)  & 0.293 & (x2.5) \\ \hline
64  & \textbf{0.086} & 2.480 & (x28.8) & 0.359 & (x4.2)  & \textbf{0.350} & 2.513 & (x7.2)  & 0.412 & (x1.2) \\ \hline
\multicolumn{11}{|c|}{} \\[-2ex]
\hline
\multicolumn{11}{|c|}{\textbf{Models D6}} \\ \hline
\multirow{2}{*}{\textbf{\begin{tabular}[c]{@{}c@{}}Time-\\Steps\end{tabular}}} & \multicolumn{5}{c||}{\textbf{\texttt{LSTM-AE-F32-D6}}} & \multicolumn{5}{c|}{\textbf{\texttt{LSTM-AE-F64-D6}}} \\
\cline{2-11} 
& \textbf{FPGA} & \multicolumn{2}{c|}{\textbf{CPU}} & \multicolumn{2}{c|}{\textbf{GPU}} & \textbf{FPGA} & \multicolumn{2}{c|}{\textbf{CPU}} & \multicolumn{2}{c|}{\textbf{GPU}} \\
\hline
 1  & \textbf{0.038} & 1.155 & (x30.4) & 0.659 & (x17.3) & \textbf{0.060} & 1.208 & (x20.1) & 0.664 & (x11.1) \\ \hline
 2  & \textbf{0.036} & 1.341 & (x37.3) & 0.655 & (x18.2) & \textbf{0.066} & 1.551 & (x23.5) & 0.663 & (x10.0) \\ \hline
 4  & \textbf{0.038} & 1.643 & (x43.2) & 0.668 & (x17.6) & \textbf{0.079} & 1.774 & (x22.5) & 0.674 & (x8.5) \\ \hline
 6  & \textbf{0.038} & 1.873 & (x49.3) & 0.671 & (x17.7) & \textbf{0.093} & 1.794 & (x19.3) & 0.672 & (x7.2) \\ \hline
16  & \textbf{0.051} & 2.620 & (x51.4) & 0.710 & (x13.9) & \textbf{0.161} & 2.697 & (x16.8) & 0.701 & (x4.4) \\ \hline
64  & \textbf{0.089} & 7.080 & (x79.6) & 0.888 & (x10.0) & \textbf{0.474} & 7.218 & (x15.2) & 0.902 & (x1.9) \\ \hline
\end{tabular}%
\vspace{-10ex}
\end{table}

\begin{table}[htbp]
\centering
\caption{Energy Per Timestep (mJ)}
\label{tab:Energy}
\resizebox{\linewidth}{!}{
\begin{tabular}{|c|c|c@{}c|c@{}c||c|c@{}c|c@{}c|}
\hline
\multicolumn{11}{|c|}{\textbf{Models D2}} \\ \hline
\multirow{2}{*}{\textbf{\begin{tabular}[c]{@{}c@{}}Time-\\Steps\end{tabular}}} & \multicolumn{5}{c||}{\textbf{\texttt{LSTM-AE-F32-D2}}} & \multicolumn{5}{c|}{\textbf{\texttt{LSTM-AE-F64-D2}}} \\
\cline{2-11} 
& \textbf{FPGA} & \multicolumn{2}{c|}{\textbf{CPU}} & \multicolumn{2}{c|}{\textbf{GPU}} & \textbf{FPGA} & \multicolumn{2}{c|}{\textbf{CPU}} & \multicolumn{2}{c|}{\textbf{GPU}} \\
\hline
 1  & \textbf{0.362} & 107.409 & (x296.7) & 9.869 & (x27.3)  & \textbf{0.435} & 108.196 & (x248.7) & 9.873 & (x22.7) \\ \hline
 2  & \textbf{0.198} & 62.321  & (x314.8) & 4.910 & (x24.8)  & \textbf{0.286} & 69.625  & (x243.4) & 4.973 & (x17.4) \\ \hline
 4  & \textbf{0.101} & 35.670  & (x353.2) & 2.430 & (x24.1)  & \textbf{0.170} & 39.853  & (x234.4) & 2.549 & (x15.0) \\ \hline
 6  & \textbf{0.071} & 25.416  & (x358.0) & 1.651 & (x23.3)  & \textbf{0.134} & 25.588  & (x191.0) & 1.703 & (x12.7) \\ \hline
16  & \textbf{0.034} & 14.538  & (x427.6) & 0.652 & (x19.2)  & \textbf{0.088} & 14.884  & (x169.1) & 0.671 & (x7.6)  \\ \hline
64  & \textbf{0.016} & 10.098  & (x631.1) & 0.204 & (x12.8)  & \textbf{0.067} & 10.111  & (x151.0) & 0.237 & (x3.5)  \\ \hline
\multicolumn{11}{|c|}{} \\[-2ex]
\hline
\multicolumn{11}{|c|}{\textbf{Models D6}} \\ \hline
\multirow{2}{*}{\textbf{\begin{tabular}[c]{@{}c@{}}Time-\\Steps\end{tabular}}} & \multicolumn{5}{c||}{\textbf{\texttt{LSTM-AE-F32-D6}}} & \multicolumn{5}{c|}{\textbf{\texttt{LSTM-AE-F64-D6}}} \\
\cline{2-11} 
& \textbf{FPGA} & \multicolumn{2}{c|}{\textbf{CPU}} & \multicolumn{2}{c|}{\textbf{GPU}} & \textbf{FPGA} & \multicolumn{2}{c|}{\textbf{CPU}} & \multicolumn{2}{c|}{\textbf{GPU}} \\
\hline
 1  & \textbf{0.426} & 305.307 & (x716.7) & 24.002 & (x56.3) & \textbf{0.677} & 320.644 & (x473.6) & 24.189 & (x35.7) \\ \hline
 2  & \textbf{0.201} & 179.089 & (x891.0) & 11.912 & (x59.3) & \textbf{0.381} & 207.116 & (x543.6) & 12.106 & (x31.8) \\ \hline
 4  & \textbf{0.107} & 109.476 & (x1023.1)& 6.080  & (x56.8) & \textbf{0.235} & 118.339 & (x503.6) & 6.170  & (x26.3) \\ \hline
 6  & \textbf{0.072} & 83.437  & (x1158.8)& 4.064  & (x56.4) & \textbf{0.185} & 79.808  & (x431.4) & 4.107  & (x22.2) \\ \hline
16  & \textbf{0.037} & 43.461  & (x1174.6)& 1.615  & (x43.6) & \textbf{0.125} & 45.089  & (x360.7) & 1.612  & (x12.9) \\ \hline
64  & \textbf{0.017} & 29.275  & (x1722.1)& 0.555  & (x32.6) & \textbf{0.097} & 30.067  & (x309.9) & 0.563  & (x5.8)  \\ \hline
\end{tabular}%
} 
\vspace{-15pt}
\end{table}


Table \ref{tab:LatencyComparison} details the average latency over 1000 inferences, for each implementation, and shows our FPGA accelerator consistently achieved lower latency. 
Speedups reached 7.2x-79.6x vs. CPU and 1.2x-18.2x vs. GPU. 
The FPGA's dedicated hardware provides inherently lower base latency, yielding advantages even at a single timestep, before temporal parallelism fully benefits from sequences being longer than the network depth.

Latency scaling with timesteps depends on $RH_m$. 
For narrower models, FPGA latency scales efficiently, remaining fastest despite a steeper relative increase than GPUs due to its much lower base latency. 
Wider models (higher $RH_m=4,8$) exhibit more serialization per timestep in the bottleneck layer, leading to less favorable scaling than $RH_m=1$ cases or GPUs. 
Still, the FPGA maintains lowest overall latency in all scenarios.

The architecture shows excellent depth scalability. Tripling layers triples CPU latency and more than doubles GPU latency (e.g., \texttt{LSTM-AE-F64-D2} vs. \texttt{LSTM-AE-F64-D6} at 64 timesteps: ~2.9x CPU, ~2.2x GPU). 
FPGA latency increases more modestly ($\sim$1.4x). 
This superior scalability stems from exploiting temporal parallelism and balancing the dataflow, where computations overlap across layers, hiding much of the added depth's latency.

Table \ref{tab:Energy} shows FPGA energy-per-timestep reductions of 151.0x-1722.1x vs. CPU, and 3.5x-59.3x vs. GPU. 
These energy savings come from low latency and significantly lower FPGA power (11-12W vs. 255-265W CPU, ~35-40W GPU). 
While energy-per-timestep decreases with sequence length for all platforms, the FPGA's low power ensures superior efficiency, combining latency benefits with inherent power advantages.


\section{Conclusion}
\label{sec:Conclusion}

We presented a novel FPGA-based LSTM-AE accelerator employing a dataflow architecture to exploit \textit{temporal parallelism} across LSTM layers. 
Our dataflow balancing methodology, using configurable hardware reuse factors, equalizes per-timestep latency, maximizing utilization and throughput.
Evaluations on four LSTM-AE models on a ZCU104 MPSoC FPGA showed significant advantages over CPU and GPU baselines.  These results validate our balanced dataflow architecture for harnessing temporal parallelism, enabling high-performance, energy-efficient LSTM-AE acceleration on FPGAs for real-time anomaly detection.


\bibliographystyle{splncs04}

\end{document}